\documentclass[twocolumn,aps,prl] {revtex4} 
\usepackage{graphicx}
\begin{document}
\pagestyle{empty} 
\title{Contact mechanics: relation between interfacial separation and load}
\author{  
B.N.J. Persson} 
\affiliation{IFF, FZ-J\"ulich, 52425 J\"ulich, Germany}

\begin{abstract}
I study the contact between a rigid solid with a randomly rough surface
and an elastic block with a flat surface. 
I derive a relation between the (average) interfacial separation $u$
and the applied normal squeezing pressure $p$. I show that for non-adhesive interaction
and small applied pressure, $p \sim {\rm exp} (-u/u_0)$, in good agreement with
recent experimental observation.
\end{abstract}
\maketitle


When two elastic solids with rough surfaces are squeezed together, the solids will
in general not make contact everywhere in the apparent contact area,
but only at a distribution of asperity contact spots\cite{Borri,Hyun,Chunyan,Carlos}. The separation
$u({\bf x})$ between the surfaces will vary in a nearly random way with the lateral
coordinates ${\bf x}=(x,y)$ in the apparent contact area. 
When the applied squeezing pressure increases, the average 
surface separation $u=\langle u({\bf x})\rangle$ will decrease, but in most situations it is not
possible to squeeze the solids into perfect contact corresponding to $u=0$. The space between
two solids has a tremendous influence on many important processes, e.g., heat transfer\cite{heat},
contact resistivity\cite{Rab}, lubrication\cite{BookP}, sealing\cite{sealing}, 
optical interference\cite{Benz}, ... . 
In this paper I will present a very simple
theory for the (average) separation $u$ as a function of the squeezing pressure $p$.
I will show that for randomly rough surfaces at low squeezing pressures $p \sim {\rm exp} (-u/u_0)$
where the reference length $u_0$ depends on the nature of the surface roughness but is independent of $p$,
in good agreement with experiments\cite{Benz}.

\begin{figure}
\includegraphics[width=0.45\textwidth,angle=0]{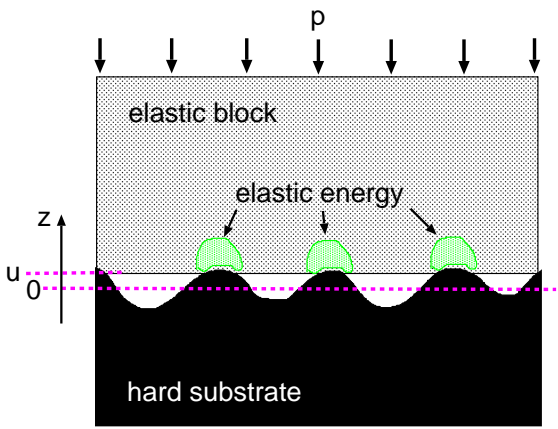}
\caption{\label{block}
An elastic block squeezed against a rigid rough substrate. The separation
between the average plane of the substrate and the average plane of the lower
surface of the block is denoted by $u$. Elastic energy is stored in the block in the vicinity
of the asperity contact regions.
}
\end{figure}

We consider the frictionless contact between elastic solids with randomly rough surfaces.
If $z=h_1({\bf x})$ and $h_2({\bf x})$ describe the surface profiles, $E_1$ and $E_2$ are
the Young's elastic moduli of the two solids and $\nu_1$ and $\nu_2$ the corresponding
Poisson ratios, then the elastic contact problem is equivalent to the contact between a
rigid solid (substrate) with the roughness profile $h({\bf x}) = h_1({\bf x})+h_2({\bf x})$, in contact
with an elastic solid (block) with a flat surface and with an Young's modulus $E$ and Poisson ratio
$\nu$ choosen so that\cite{Johnson2}
$${1-\nu^2 \over E} = 
{1-\nu_1^2 \over E_1} + 
{1-\nu_2^2 \over E_2}.$$ 

Introduce a coordinate system $xyz$ with the $xy$-plane in the average
surface plane of the rough substrate, and the $z$-axis pointing
away from the substrate, see Fig. \ref{block}. 
The separation between the average surface plane of the block and the
average surface plane of the substrate is denoted by $u$ with $u \ge 0$.
When the applied squeezing force $p$
increases, the separation between the surfaces at the interface will decrease, and
we can consider $p=p(u)$ as a function of $u$. 
The elastic energy $U_{\rm el}(u)$ stored in the substrate asperity--elastic block contact regions
must equal to the work done by the external pressure $p$ in displacing the lower surface of the
block towards the
substrate, i.e.,
$$\int_u^\infty du' A_0 p(u') = U_{\rm el}(u)\eqno(1)$$ 
or
$$p(u) = - {1\over A_0} {d U_{\rm el} \over du},\eqno(2)$$ 
where $A_0$ is the nominal contact area. Equation (2) is exact.
Theory shows that for low squeezing pressure, the area of real contact $A$ varies linearly with
the squeezing force $pA_0$, and that the interfacial stress distribution, and the 
size-distribution of contact spots, are independent of the squeezing pressure\cite{Arch,PSSR}. 
That is, with increasing $p$
existing contact areas grow and new contact areas form in such a way that in the thermodynamic limit
(infinite-sized system) the quantities referred to above remain unchanged. It follows immediately
that for small load {\it the elastic energy stored in the asperity contact region will increase
linearly with the load}, i.e., $U_{\rm el}(u) = u_0 A_0 p(u)$, where $u_0$ is a characteristic length
which depends on the surface roughness
(see below) but is independent of the squeezing pressure $p$. Thus, for small pressures (2) takes the form
$$p(u) = - u_0 {d p \over du}$$
or
$$p(u) \sim e^{- u/u_0}\eqno(3)$$ 
in good agreement with experimental data for the contact between elastic solids when the
adhesional interaction between the solids can be neglected\cite{Benz}. 
We note that the result (3) differs drastically from
the prediction of the Bush et al theory\cite{Bush}, and the theory of Greenwood and Williamson (GW)\cite{GW}, 
which for low squeezing pressures (for randomly rough
surfaces with Gaussian height distribution) predict $p(u) \sim u^{-a} {\rm exp} (-bu^2)$, where $a=1$ in the
Bush et al theory and $a=5/2$ in the GW theory. Thus {\it these theories do not correctly describe the
interfacial spacing between contacting solids}.

\begin{figure}
\includegraphics[width=0.45\textwidth,angle=0]{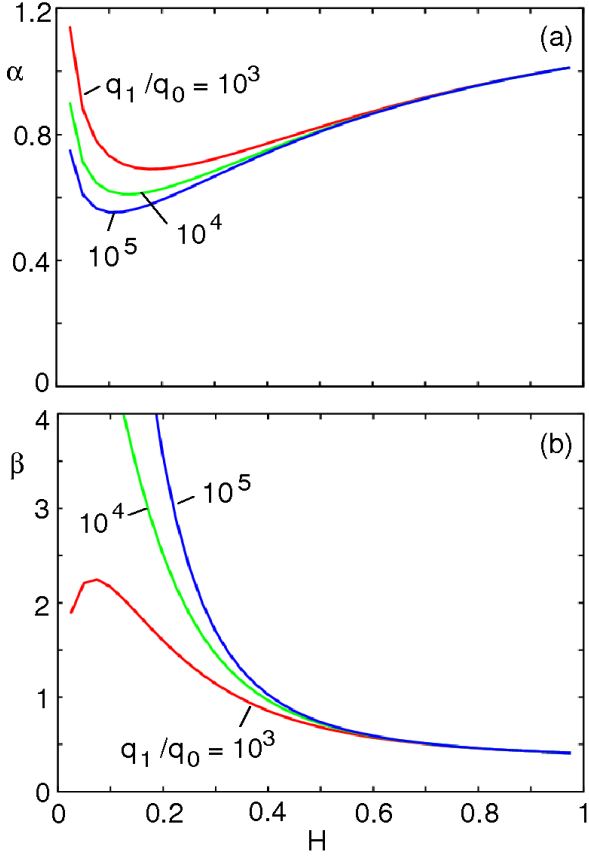}
\caption{\label{alfbet}
The parameters $\alpha$ and $\beta$ as a function of the Hurst exponent
$H$ for three different values of the ratio $q_1/q_0$. 
}
\end{figure}

The elastic energy $U_{\rm el}$ has been studied in Ref. \cite{P1} and \cite{PSSR}, 
and in the simplest approximation it takes the form
$$U_{\rm el} \approx A_0 E^* {\pi \over 2}\gamma \int_{q_0}^{q_1} dq \ q^2P(q)C(q),\eqno(4)$$
where $E^*=E/(1-\nu^2)$ and
where $P(q)= A(\zeta) /A_0$ is the relative contact area when the interface is studied at
the magnification $\zeta = q/q_0$. The surface roughness power spectrum\cite{P3}
$$C(q)= {1\over (2\pi )^2} \int d^2x \langle h({\bf x}h({\bf 0})\rangle
e^{-i{\bf q}\cdot {\bf x}},$$
where $\langle..\rangle$ stands for ensemble average.  
The parameter $\gamma = 1$ in the simplest case but in general one expect $\gamma < 1$
(but of order unity) to take into account that the elastic energy stored in the contact
region (per unit surface area) is less than the 
average elastic energy (per unit surface area) for perfect contact,
see Ref. \cite{PSSR}. 
We will use the contact mechanics theory of Persson,
where for elastic non-adhesive contact the function\cite{JCPpers,Bucher} 
$$P(q) = {2\over \surd \pi} \int_0^{s(q)p} dx \ e^{-x^2}\eqno(5)$$
where $s(q)=w(q)/E^*$ with 
$$w(q)=\left (\pi \int_{q_0}^q dq' \ q'^3 C(q') \right )^{-1/2}\eqno(6)$$
Using (5) gives
$${\partial P \over \partial u} = {2\over \surd \pi} s {d p \over d u} e^{-s^2p^2}\eqno(7)$$
Substituting (4) and (7) in (2) gives
$$p(u)=-\surd \pi \gamma \int_{q_0}^{q_1} dq \ q^2C(q) w(q) e^{-[w(q) p/E^*]^2} {d p \over d u} $$ 
or
$$du =-\surd \pi \gamma \int_{q_0}^{q_1} dq \ q^2C(q) w(q) e^{-[w(q) p/E^*]^2} {d p \over p} $$
Integrating this from $u=0$ (complete contact, corresponding to $p=\infty$) to $u$ gives
$$u= \surd \pi \gamma \int_{q_0}^{q_1} dq \ q^2C(q) 
w(q) \int_p^\infty dp' \ {1 \over p'} e^{-[w(q) p'/E^*]^2}\eqno(8)$$

For very small pressures we get from (8):
$$u= - u_0 
{\rm log} (p/p_c)$$
or
$$p=p_c e^{- u/u_0},\eqno(9)$$
where 
$$u_0 = \surd \pi \gamma \int_{q_0}^{q_1} dq \ q^2 C(q) w(q)\eqno(10)$$
and where the cut-off
$p_c$ is determined by
$$p_c= E^* {\rm exp} (-\langle {\rm log} w\rangle)\eqno(11)$$
with  
$$\langle {\rm log} w\rangle = {\int_{q_0}^{q_1} dq \ q^2C(q) w(q) {\rm log} w(q)\over
\int_{q_0}^{q_1} dq \ q^2C(q) w(q)}\eqno(12)$$
If we assume that the substrate surface roughness is self affine fractal for $q_0 < q < q_1$ we get\cite{P3}
$$C(q) = {H\over \pi} {\langle h^2 \rangle \over q_0^2} \left ({q\over q_0}\right )^{-2(H+1)}\eqno(13)$$
where $H=3-D_{\rm f}$, where $D_{\rm f}$ is the fractal dimension. 
The mean of the square of the substrate surface height profile is $\langle h^2 \rangle = h_{\rm rms}^2$.
Substituting (13) in (10) gives
$$u_0 = h_{\rm rms}/ \alpha\eqno(14)$$ 
where
$$\alpha^{-1} = \left ({2H(1-H)\over \pi}\right )^{1/2}
\gamma \int_1^{q_1/q_0} dx \ g(x) \eqno(15)$$
with
$$g(x)=x^{-H} \left (x^2-x^{2H} \right )^{-1/2}$$
Substituting (13) in (12) gives
$$\langle {\rm log} w\rangle = - {\rm log} (q_0 h_{\rm rms})-{\rm log} \beta\eqno(16)$$
where
$${\rm log} \beta =
{\int_1^{q_1/q_0} dx \ g(x) {\rm log}
\left [ {H\over 2(1-H)}\left (x^{2(1-H)}-1\right )  \right ] \over
2 \int_1^{q_1/q_0} dx \ g(x)}.\eqno(17)$$ 
Substituting (14), (16) and (11) in (9) gives
$$p={\beta q_0 h_{\rm rms} E^*} e^{-\alpha u/h_{\rm rms}}\eqno(18)$$
or
$$u=\alpha^{-1} h_{\rm rms} {\rm log} \left ({\beta q_0h_{\rm rms} E^*/p}\right )\eqno(19)$$
In Fig. \ref{alfbet} we show 
the parameters $\alpha$ (for $\gamma = 1$) and $\beta$ as a function of the Hurst exponent
$H$ for three different values of the ratio $q_1/q_0$. 
Most surfaces which are self affine fractal have the Hurst exponent $H > 0.5$
(or the fractal dimension $D_{\rm f} < 2.5$). For such surfaces the parameters $\alpha$ and $\beta$ 
are nearly independent on the ratio $q_1/q_0$ between the highest $q_1$ and smallest $q_0$
wavevector included in the analysis. 
Surfaces prepared by crack propagation or by bombardment with particles typically have $H\approx 0.8$
(or the fractal dimension $D_{\rm f} = 3-H \approx 2.2$) for which case $\alpha \approx 1 $ and 
$\beta \approx 0.5$. In this case we have
$$p\approx {0.5 q_0 h_{\rm rms} E^*} e^{- u/\gamma h_{\rm rms}}\eqno(20)$$
or
$$u \approx \gamma h_{\rm rms} {\rm log}(0.5 q_0h_{\rm rms} E^*/p)\eqno(21)$$
where we have reintroduced the factor of $\gamma$.

\begin{figure}
\includegraphics[width=0.45\textwidth,angle=0]{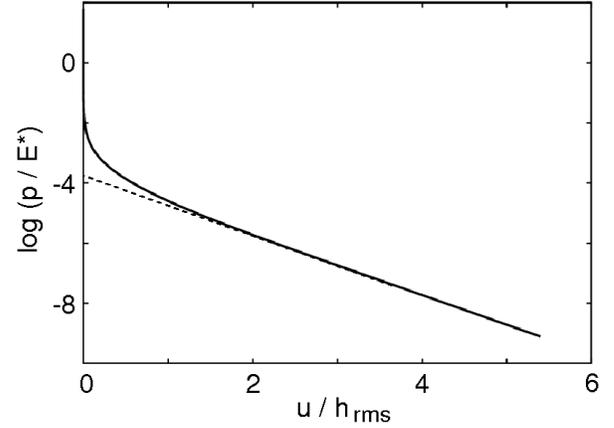}
\caption{\label{u.logp}
The relation between the (natural) logarithm of the
squeezing pressure $p$ and the interfacial separation $u$
for an elastic solid squeezed against a rigid, self affine fractal surface with
the Hurst exponent $H=0.8$. The surface has the rms roughness $h_{\rm rms} = 6 \ {\rm \mu m}$
and the upper and lower cutoff wavevectors are $q_1=7.8 \times 10^9 \ {\rm m}^{-1}$ 
and $q_0=1\times 10^4 \ {\rm m}^{-1}$. 
}
\end{figure}

In Fig. \ref{u.logp} we show the relation (for $\gamma = 1$)
between the (natural) logarithm  of the squeezing 
pressure $p$ and the interfacial separation $u$
for an elastic solid squeezed against a rigid, self affine fractal surface with
the Hurst exponent $H=0.8$, as calculated directly from (8) by numerical integration. 
The dashed line shows the large-distance asymptotic behavior
given by (21). Note that deviation from this logarithmic
relation only occurs for $u < h_{\rm rms}$,
and that there is a sharp increase in the squeezing pressure for $u < h_{\rm rms}$.
Both these facts are in 
accordance with the experimental data presented in Ref. \cite{Benz}.
Note that the slope of the dashed line in Fig. \ref{u.logp} is close to $-1$,
as expected from (21), and that the dashed curve for $u=0$ gives ${\rm log}(p/E^*) \approx -3.8$
or $p/E^* \approx 0.02$, which is similar to the prediction 
of (21): $0.5 q_0h_{\rm rms} \approx 0.03$. I have performed calculations of the $p(u)$
relation using several measured surface roughness power spectra $C(q)$, but the general form
is always as in Fig. \ref{u.logp}, and the slope of the line (using $\gamma = 1$) in the linear
region is always close to $-1$. 

Pei et al\cite{Pei} have performed a finite element method 
computer simulation of the contact mechanics
for a polymer surface, using the measured
surface topography\cite{Benz} as input, squeezed against a flat surface. They found that
for large separation $p\sim {\rm exp} (-u/\gamma h_{\rm rms})$ with $\gamma \approx 0.6$,
which is consistent with our numerical and analytical results. 

The theory presented above can be easily generalized in various ways.
Thus, it is possible to include the adhesional interaction\cite{Full,Johnson1}.
In this case the work done by the external pressure $p$ will be the sum of the stored (asperity
induced) elastic energy plus the (negative) adhesional energy, i.e., the right hand side of
(1) will now be $U_{\rm el}+U_{\rm ad}$. 
The theory can also be 
applied to study how the spacing $u(\zeta)$ depends on the magnification. Here 
$u(\zeta)$ is the (average) spacing between the solids in the apparent contact areas
observed at the magnification $\zeta$. The quantity $u(\zeta)$ is of crucial importance for
lubricated seals\cite{preparation}. 
The results of these generalizations of the theory will 
be presented elsewhere.

Finally we note that the observation of an effective exponential repulsion has important 
implications for tribology, colloid science,
powder technology, and materials science\cite{Benz}. 
For example, the density or volume of granular materials has long been known to have a
logarithmic dependence on the externally applied isotropic pressure or stress, as found, for example,
in the compression stage during processing of ceramic materials\cite{Stanley}. Recent work on the confinement of
nanoparticles have also indicated an exponential force upon compression\cite{Alig},
suggesting that this relationship could be prevalent among quite different 
types of heterogeneous surfaces.

\end{document}